\documentstyle[aps,epsfig,twocolumn,citesort,psfig19,floats]{revtex}
\renewcommand {\baselinestretch}{1.0}

\def\begineq{\begin{equation}}
\def\endeq{\end{equation}}
\def\be{\begin{equation}}
\def\ee{\end{equation}}

\begin{document}
\bibliographystyle{prsty}
%\psdraft

\title{
Water temperature dependence of single bubble sonoluminescence
}
\author{
Sascha Hilgenfeldt $^1$,
Detlef Lohse $^1$, and 
William C. Moss $^2$
}
\address{
$^1$
Fachbereich Physik der Universit\"at Marburg,
Renthof 6, D-35032 Marburg\\
$^2$ Lawrence Livermore National Laboratory, Livermore, CA 94550
}

\date{Phys. Rev. Lett., February 16, 1998}

\maketitle
\begin{abstract}
The strong dependence of the intensity of single bubble
sonoluminescence 
(SBSL) on water temperature observed in experiment can
be accounted for by the temperature dependence of the
material constants of water, most essentially
of the viscosity,
of the argon solubility in water,
and of the 
vapor pressure.
The strong increase of light emission at low water temperatures
is due to the possibility of applying higher driving pressures,
caused by increased bubble stability.  
The presented calculations combine the Rayleigh-Plesset equation based
hydrodynamical/chemical approach to SBSL and full gas dynamical
calculations of the bubble's interior. 
\end{abstract}
%Pacs: 47.27.Gs

\vspace{1cm}

%\newpage

%----------------------------------------------------------------------

\begin{figure}[htb]
\setlength{\unitlength}{1.0cm}
\begin{picture}(6,6.6)
%\put(-0.5,6.5)
\put(-0.7,-0.3)
{\psfig{figure=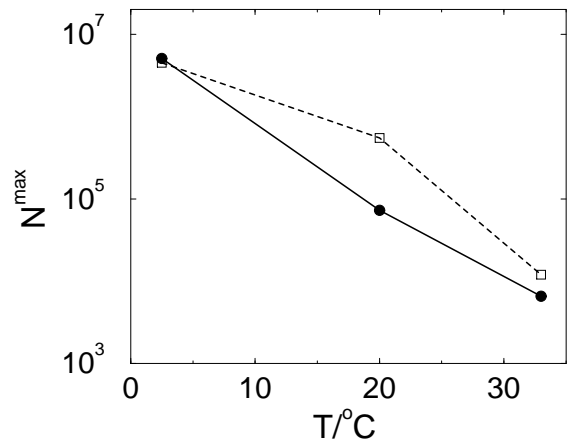,width=9cm,angle=-90}}
\end{picture}
\caption[]{
Maximally achievable number $N^{max}$ of SBSL photons for 
given water temperature. The experimental data (open boxes) are taken from
figure 1 of \cite{bar94}.
The theoretical data (filled circles) 
result from the full gas dynamical calculations
performed with the algorithm described in \cite{mos97}.
}
\label{intens}
\end{figure}

\begin{figure}[htb]
\setlength{\unitlength}{1.0cm}
\begin{picture}(6,6.6)
%\put(-0.5,6.5)
\put(-0.7,-0.3)
{\psfig{figure=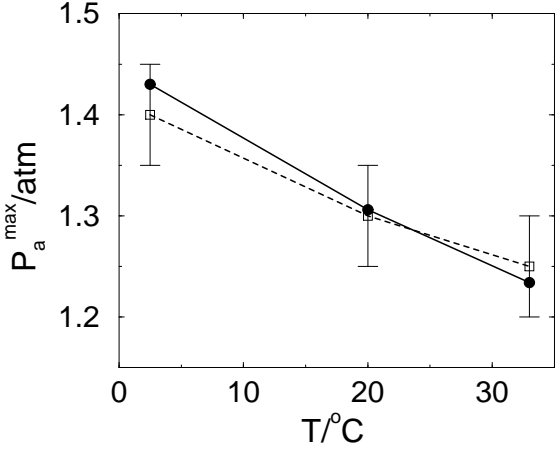,width=9cm,angle=-90}}
\end{picture}
\caption[]{
Maximum forcing pressure for stable SBSL as a
function of the water temperature. 
The experimental data (open boxes) are from
figure 1 of \cite{bar94} . The filled circles are the
theoretical values following from figure \ref{phase_dia}.  
}
\label{pa}
\end{figure}

One of the remarkable features of single bubble sonoluminescence
 (SBSL) \cite{gai90,bar97} is the
sensitivity of the light emission 
to the water temperature experimentally found by the UCLA
group \cite{bar94,bar97}, cf.\ figure \ref{intens}. 
To obtain these results, Barber et al.\
proceeded as follows (refs.\ \cite{bar97,bar94} and B.\ Barber,
private communication, August 1997): Water was cooled to
a temperature of $2.5^oC$ and completely degassed. 
An air pressure overhead of $150$ Torr, corresponding to 
about $20\%$  of gas saturation, 
was adjusted and SL
experiments were performed, still at $2.5^oC$. Then the water
was heated to $20^oC$ {\it without} readjusting the gas
concentration, and
the SL experiment was repeated. 
Finally, the same measurement was performed after heating the water 
to $33^oC$. 
At all three temperatures, the forcing pressure amplitude $P_a$ of the
driving
sound field was adjusted in order to give maximum light intensity,
while maintaining bubble stability against fragmentation ({\em stable}
 SL). 
According to the ``waterfall plots'' shown e.g.\ in  ref.\ \cite{bar97},
the highest light intensity always corresponds to the
largest achievable driving pressure ($P_a^{max}$). 
The experimental values for
$P_a^{max}$ are shown in figure \ref{pa}.

The bubble radius dynamics $R(t)$ was  detected in these
experiments (using Mie scattering techniques) and
fitted by the Rayleigh-Plesset (RP) equation \cite{bar97}. In this fit,
several parameters were allowed to vary: the driving pressure
amplitude $P_a$, the ambient radius $R_0$ (radius of the bubble under 
normal conditions) and also the surface tension $\sigma$ of the air/water
interface and the water viscosity $\nu_l$ \cite{bar94,bar97}.
We consider the treatment of the latter two quantities as free
parameters arbitrary, as $\sigma$ and $\nu_l$ are well defined
material constants, but are varied in \cite{bar94,bar97} by as much as
a factor of four and beyond. 
Fitting $\sigma$ and $\nu_l$ may  be avoided if a
more realistic model for the thermal behavior of the gas inside the
bubble is applied: in the RP equation, the internal gas pressure is
taken to vary polytropically with volume, with an effective exponent 
$\gamma$. In \cite{bar94,bar97} $\gamma$ changes abruptly from its
isothermal to its adiabatic value at $R(t)=R_0$. Taking instead
a more realistic model  $\gamma = 1$ throughout the whole
oscillation with the exception of the immediate vicinity of the strong
bubble collapse results in a
satisfactory fit with the physical values for $\sigma$ and $\nu_l$.

\begin{figure}[htb]
\setlength{\unitlength}{1.0cm}
\begin{picture}(6,6.6)
%\put(-0.5,6.5)
\put(0.2,0.5)
{\psfig{figure=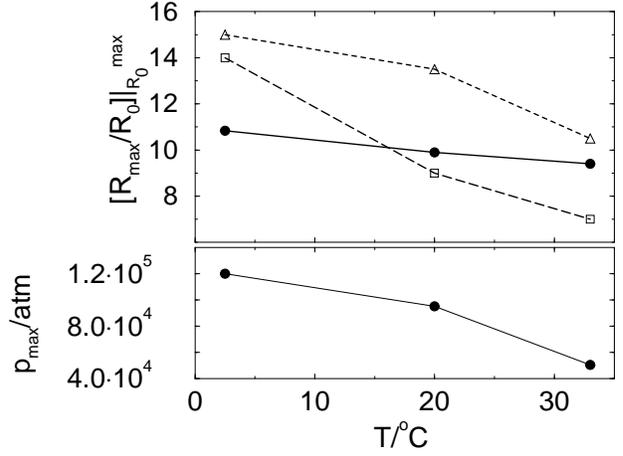,width=8cm,angle=-90}}
\end{picture}
\caption[]{
Experimental data 
for the expansion ratio
$R_{max}/R_0$ 
at maximal ambient radius $R_0^{max}$
as a function of the water temperature
taken from
figure 1 of \cite{bar94} (open triangles) 
and figure 47 of \cite{bar97} (open boxes). 
Also shown are the theoretical values following within the presented
theoretical approach (filled circles).
The lower part of the figure shows the maximal pressure achieved in the
bubble within the hydrodynamical/chemical approach. 
}
\label{rmax}
\end{figure}

Because of the complications in the fits of refs.\
\cite{bar94,bar97}, the resulting data should be read with some
care. This is also reflected in figure \ref{rmax}, where  we display  
the data from these two references for the expansion ratio (maximum
radius $R_{max}$ divided by $R_0$): they show large deviations 
at otherwise unchanged parameters. The expansion ratio is a quantity
closely related to the violence of collapse and therefore, presumably,
to the intensity of energy concentration and light emission \cite{mos97}.
It is therefore puzzling that 
the same light intensity has been observed in refs.\
\cite{bar97,bar94}
in spite of the different expansion ratios reported in figure
\ref{rmax}.

 \begin{table}[htp]
 \begin{center}
 \begin{tabular}{|c|c|c|c|}
 \hline
          water temperature
       & $2.5^oC$
       & $20^oC$
       & $33^oC$
\\
\hline
         $\sigma [kg/s^2]$  
       & $0.0753 $
       & $0.0728 $
       & $0.0707 $
\\
         $\rho [g/cm^3]$  
       & $1.000 $
       & $0.998 $
       & $0.995 $
\\
         $\nu_l /10^{-6} [m^2/s]$  
       & $1.66 $
       & $1.01 $
       & $0.75 $
\\
         $p_{vap} [kPa]$  
       & $0.734 $
       & $2.339 $
       & $5.034 $
\\
         $c_l [m/s]$  
       & $1414 $
       & $1483 $
       & $1516 $
\\
         $c_0^{Ar} [kg/m^3]$  
       & $0.0892 $
       & $0.0611 $
       & $0.0495 $
\\
         $c_0^{N_2} [kg/m^3]$  
       & $0.0399 $
       & $0.0283 $
       & $0.0238 $
\\
 \hline
 \end{tabular}
 \end{center}
\caption[]{
Material constants of water
as a function of temperature 
\cite{lid91}. From top to bottom: 
surface tension $\sigma$, density $\rho$, kinematic viscosity $\nu_l$,
vapor pressure $p_{vap}$, speed of sound
$c_l$, and solubilities of argon and nitrogen in water
$c_0^{Ar}$, $c_0^{N_2}$.
}
\label{tab_mat}
 \end{table}

The central claim of this paper is that the observed dependence on
water temperature $T$
in figures \ref{intens} -- \ref{rmax} can be accounted for by the 
$T$ dependence of the material constants of water which are listed in
table \ref{tab_mat}. In our analysis the most relevant effects are the
temperature dependence of the 
gas solubilities (as already conjectured in ref.\ \cite{hil92})
and of the viscosity; also, the temperature dependence of water vapor
contributes. 
The variations of these material constants with temperature conspire
to allow for larger shape stable bubbles and larger forcing at lower
temperatures, resulting in more light emission.

When dissolving air with its various constituents in water, 
only the
argon concentration 
$c_\infty^{Ar}$ 
(far away from the bubble)
is relevant in the SL regime,
as SL air bubbles rectify argon \cite{loh97}.
The degree of saturation $c_\infty^{Ar}/c_0^{Ar}$ is one of the central
parameters to determine the diffusive equilibrium (ambient) radius of SL
bubbles.
This equilibrium radius can be calculated within the
hydrodynamical/chemical approach to SL \cite{hil96,loh97,bre97}, which
takes into account the mass exchange between the bubble interior and
the liquid due to (i) diffusion and (ii) dissociation of molecular
gases, which leads to argon rectification in the bubble.
Here we follow that approach and calculate the phase diagrams in
$R_0-P_a$
space, resulting from the diffusional/chemical stability of the bubble for
varying temperature, i.e.,  varying material constants and also 
varying $c_\infty^{Ar}/c_0^{Ar}(T)$.
The calculations (based on Rayleigh-Plesset dynamics)
are described in detail in refs.\
\cite{loh97}.
We choose the material constants 
given in table \ref{tab_mat}, the frequency 
$26.5kHz$ as in experiment \cite{bar94}, and an ambient pressure of 1\,atm. 
We modeled air as a mixture of $99\%$ nitrogen and 
$1\%$ argon (in the gas phase above the liquid). 
At a given total gas concentration of $c_\infty/c_0 = 20\%$ for
$T=2.5^\circ C$, the corresponding argon concentrations for the three
temperatures are readily calculated from table \ref{tab_mat}:
$c_\infty^{Ar}/c_0^{Ar}(2.5^\circ C)=0.20\%$,
$c_\infty^{Ar}/c_0^{Ar}(20^oC)=0.29\%$, and 
$c_\infty^{Ar}/c_0^{Ar}(33^oC)=0.36\%$, respectively.

%The result for 
%$2.5^oC$ ($c_\infty /c_0 = 20\%)$,
%$20^oC$ ($c_\infty /c_0 = 29\%)$, and
%$33^oC$ ($c_\infty /c_0 = 36\%)$ is shown in 

\begin{figure}[htb]
\setlength{\unitlength}{1.0cm}
\begin{picture}(6,7.5)
%\put(-0.5,7.5)
\put(-0.5,0.0)
{\psfig{figure=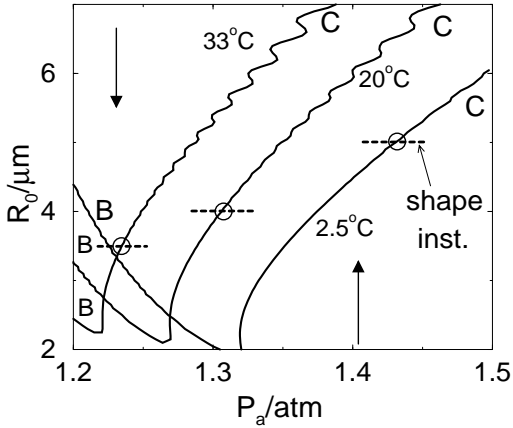,width=8cm,angle=-90}}
\end{picture}
\caption[]{
Phase diagram in the $R_0-P_a$ parameter space. 
The solid lines show the stable diffusive equilibria for three
different temperatures. The bubbles along the equilibria C
contain essentially pure argon. The bubbles along the equilibria B
have a slightly enhanced argon concentration -- these bubbles are in
equilibrium because the mass gain by rectified diffusion and the mass
loss by chemical reactions balance \cite{loh97}.
Left of each curve C bubbles are shrinking, right of each curve C they are
growing, as indicated by the arrows. 
 -- The dashed lines
(shape inst.) show where the bubbles become spherically unstable 
because of the
parametric shape instability \cite{hil96}.  
The intersection with the diffusive stability line
defines $P_a^{max}$ and $R_0^{max}$ (circles).
}
\label{phase_dia}
\end{figure}

The resulting equilibrium curves in phase space are shown in figure 
\ref{phase_dia}.
Diffusively stable bubbles are possible on the branches  B 
and C
\cite{loh97}; however, according to the
(qualitative) energy focusing condition
$|\dot R|/c_g > 1$ \cite{bre95,hil96,bar97} 
(where $c_g$ is the speed of sound in the gas bubble)
most bubbles on branch B
will not be able to emit SL light,
in agreement with recent experiments \cite{hol96,bre97}. 

A key issue is that the upper limit in both $P_a$ and $R_0$ on
branch C is given by the parametric shape instability of the bubble wall
\cite{ple54,bre95,hil96,comment}.
The shape instability depends on the {\it viscosity} 
of the water \cite{hil97}, which
strongly increases with decreasing water temperature,
thus stabilizing the bubble. 
For the values given in table \ref{tab_mat}
 we find (within the approximations
of \cite{hil96}) that shape instability sets in at an ambient radius
of roughly $R_0=3.5\mu m$, $4\mu m$, and $5\mu m$ for 
$33^oC$,
$20^oC$, and
$2.5^oC$, respectively, see the dashed lines in figure
\ref{phase_dia}.

{}From figure \ref{phase_dia} one immediately realizes that the bubble
can be driven harder (larger $P_a^{max}$) for lower 
temperatures.
The values
$P_a^{max}$ from figure \ref{phase_dia} are displayed in figure \ref{pa}, 
showing good agreement with experiment. 

{\it On decreasing the water temperature},
the material constants of water are
changed according to table \ref{tab_mat}; this has the following effect on  
the curves in figure \ref{phase_dia}:
(i) 
The {\em water viscosity} $\nu_l$ is {\it increased}; as
stated above, $\nu_l$ determines the damping of bubble shape oscillations;
therefore, bubbles are more stable; the stability line shifts towards larger
$R_0$.
(ii) 
The {\em relative
  argon concentration} $c_\infty^{Ar}/c_0^{Ar}(T)$ becomes {\it smaller};
  in order
  to counteract dissolution, the bubbles must oscillate more violently to
  rectify gases. Thus, diffusive equilibria are reached
  only at higher $P_a$, 
and curves B and C are shifted to the right, allowing for
larger
$P_a^{max}$.
(iii) 
The {\em water vapor pressure} $p_{vap}$ {\it decreases};
this, too, slightly shifts curves B and C to 
the right. The reason is that
the reduced total pressure inside the bubble leads to smaller
$R_{max}$. The resulting reduced gas rectification has to be compensated
by larger $P_a$. 
(iv)
The small temperature
dependences of $\sigma$, $\rho$, and $c_l$ have hardly any effect.

In figure \ref{rmax} we include
the theoretical expansion ratio
$R_{max}/R_0|_{R_0^{max}}$ calculated 
with the Rayleigh-Plesset equation for $P_a^{max}$ and
$R_0^{max}$ 
determined from figure \ref{phase_dia}.
Here, the agreement with the data from the 
fits of \cite{bar94,bar97} is not as good. 
We speculate that part of
the discrepancy originates from the fit procedure
(to obtain $R_0$) of refs.\ \cite{bar94,bar97} (see our criticism
above) and suggest direct measurements of $R_{max} / R_0$, either by
Mie scattering at two different frequencies or by employing the method
developed by Holt and Gaitan \cite{hol96}.

One may wonder why a relatively modest change in $P_a$ and $R_0^{max}$
(cf.\ figs.\ \ref{phase_dia} and \ref{pa}) leads to considerably enhanced
light emission, cf.\ fig.\ \ref{intens}.
The reason is twofold: (i) With increasing $P_a$ the bubble
collapse becomes {\it much} more violent and therefore 
the gas inside the bubble undergoes stronger heating.
The violence of the bubble collapse
can be quantified by maximal pressure $p_{max}$ reached
inside the bubble (within the hydrodynamical approach), cf.\ fig.\
\ref{rmax}, lower part.
The observed increase
with decreasing water temperature is a direct consequence
of the observed {\it increase} in $R_{max}/R_0$ (upper part of fig.\
\ref{rmax}): For bubbles with a larger expansion ratio, more
potential energy can be converted into kinetic energy during the collapse,
compressing the bubble more strongly.
%during the bubble extension
%is eventually transfered into heat at the bubble collapse. 
(ii) Larger $R_0^{max}$ means that a larger number of gas molecules
$\propto (R_0^{max})^3$ can be heated, which also leads to more light
emission. 

To complete our analysis, we would like to
{\it calculate} the light
intensity and compare it
with the experimental data in figure \ref{intens}. 
The light intensity can of course not be obtained from the
RP approach to SBSL. 
It sensitively depends on the exact temperature achieved in
the bubble at the collapse which can only be {\it estimated} within the
Rayleigh-Plesset bubble approach, using one of a number of simple
models, e.g., adiabatic heating at the collapse. 

However, numerical codes that calculate the gas dynamics of the bubble's
{\it interior} and the resulting optical emission have been developed
\cite{wu93,mos97}.
We use
the model recently proposed by Moss et al.\ \cite{mos97}
to calculate the gas dynamics inside the bubble at the collapse
and the resulting optical emission.  The model solves the gas dynamic
equations for the conservation of mass, momentum, and energy in the gas
bubble and the surrounding liquid, using a finite differencing scheme for
the assumed spherical growth and collapse of the bubble.  Accurate high
pressure/temperature equations of state are used to describe the water and
the argon.  Energy loss by thermal conduction in the
partially ionized gas created during
the collapse is included in the model. Although light emission is the
primary diagnostic of SBSL, the energy loss by the radiation of light is
negligible compared to that by thermal conduction. This simplifies
the calculation of the emitted optical power and spectra, which  are
calculated using the opacity and the calculated densities and electron
temperatures of the partially ionized gas.
Full details are given in \cite{mos97}.
The model
was able to account for many experimentally found features of
SBSL, 
including spectra and the sensitivity of the
spectral intensity to
the applied acoustic pressure (or, equivalently,
$R_{max}$).
This last result establishes the relevance of the model to the
current analysis.
We use the parameter pairs 
($P_a^{max}$, $R_0^{max}$) or equivalently 
($R_{max}|_{R_0^{max}}$, $R_0^{max}$)
resulting from above hydrodynamical/chemical approach 
for the three different analyzed water temperatures
as input parameters for the gas dynamical code. 
The number of photons obtained in the wavelength window $[200nm,
750nm]$ is compared to the experimental data in figure \ref{intens},
showing reasonable agreement.

To summarize, the experimentally found water temperature dependence
of SBSL \cite{bar97,bar94}
can consistently be accounted for by the water temperature dependence
of the material constants of water: At lower water temperature
bubbles are more stable and can therefore
be forced more strongly,
resulting in more light. Combined with a gas dynamical simulation of the
bubble's interior, this approach can reproduce the three orders of magnitude
increase in the observed photon number when changing the temperature from
$\sim 30^oC$ to $\sim 0^oC$.

We suggest several control experiments to study the
temperature dependence of SBSL. In all cases the experiments should be
done with {\it stable} SBSL, whose regime in phase space can always 
(at least approximately) be
calculated  a priori, following the hydrodynamical/chemical approach. 

(i) The bubble dynamics $R(t)$ should be measured as a function of the
water temperature -- this should reveal that the water temperature
dependence of SBSL is mainly a bubble dynamical effect. In particular,
as already mentioned above, the experimental data of figure \ref{rmax}
should be remeasured. 

(ii) In addition, as suggested in a recent preprint by Vuong, Fyrillas,
and Szeri \cite{szeri}, one should study how the bubble dynamics and
the light intensity change (for fixed forcing pressure) with water
temperature for {\it xenon doped} and {\it helium doped} nitrogen
bubbles. The two noble gases should show different behavior, because
$c_0^{Xe}$ strongly depends on the water
temperature $T$, whereas $c_0^{He}$ only shows a weak dependence, see
figure 3 of ref.\ \cite{szeri}. This kind of experiment will be a
useful diagnostics of the identity of the gas in the bubble \cite{szeri}.
It will also help to distinguish between 
the gas solubility effects on the light intensity and other bubble
dynamical effects, e.g., those caused by the temperature dependence of
the water viscosity and of the vapor pressure.

(iii) Finally, instead of preparing the
gas concentration at one temperature
(resulting in a water temperature dependent {\it relative}
concentration
$c_\infty^{Ar}/c_0^{Ar}(T)$), one could {\it keep}
$c_\infty^{Ar}/c_0^{Ar}$
{\it constant}
 at various temperatures, e.g.\ by using noble gas doped
{ oxygen} bubbles and 
{ controlling} the actual gas content with oxyometry, cf.\
Gompf et al.\ \cite{gom97}.
If in addition the forcing pressure amplitude is kept constant, 
the location of the shape instability and its dependence on water
viscosity will also become irrelevant.
Altogether, we therefore predict that the water 
temperature dependence of SBSL light emission {\it at otherwise  fixed
parameters} $P_a$, $c_\infty^{Ar}/c_0^{Ar}$ 
will be only weak.

\vspace{0.3cm}

\noindent
{\bf Acknowledgments:}
We thank all participants of the NATO-ASI workshop on Sonoluminescence
and Sonochemistry for various discussions and in particular Larry Crum
for the organization of that workshop. 
We are very grateful to M.\ Brenner and S.\ Grossmann 
for invaluable discussions and stimulating comments.
Support for this work by
the DFG under grant SBF185-D8 and by a joint NSF/DAAD grant
is acknowledged. WCM's work was performed under the auspices of the U.S.
Department of Energy by LLNL under contract W-7405-Eng-48.
S. H. and D. L. thank L.\ Kadanoff and the University of Chicago
(where part of this work has been done)
for hospitality and acknowledge partial support by
the Office of Naval
Research.

%\noindent
% e-mail addresses:\\
%lohse@stat.physik.uni-marburg.de\\

\vspace{-0.5cm}

%\bibliography{sl_literatur}

\begin{thebibliography}{10}
\bibitem{comment}
The theoretical results for the onset of shape instabilities
seem to (slightly)
underestimate the experimental ones \cite{hol96}, possibly because
heat conduction was neglected. 
For a detailed
comparison
between the theoretical and experimentally found onset of shape
instabilities we refer the reader to reference
\cite{bre97}.
\end{thebibliography}

\end{document}